\documentclass[aps,twocolumn,english,superscriptaddress,citeautoscript,preprintnumbers,amsmath,amssymb,floatfix,footinbib]{revtex4-1}
\usepackage[breaklinks=true,colorlinks,citecolor=blue,linkcolor=blue,urlcolor=blue]{hyperref}
\usepackage{amsmath}
\usepackage{graphicx}
\usepackage{dcolumn}
\usepackage{float}
\usepackage[utf8]{inputenc}
\usepackage{color}
\usepackage[utf8]{inputenc}
\usepackage[T1]{fontenc}
\begin{document}
	
\title{ Large magnetoresistance and first-order phase transition in antiferromagnetic single-crystalline EuAg$_4$Sb$_2$}
\author{Sudip Malick}
\email{sudip.malick@pg.edu.pl}
\author{Hanna Świątek}
\affiliation{Faculty of Applied Physics and Mathematics, Gdansk University of Technology, Narutowicza 11/12, 80-233 Gdańsk, Poland}
\affiliation{Advanced Materials Center, Gdansk University of Technology, Narutowicza 11/12, 80-233 Gdańsk, Poland}
\author{Joanna Bławat}
\author{John Singleton}
\affiliation {National High Magnetic Field Laboratory, Los Alamos National Laboratory,
Los Alamos, NM 87545, USA}
\author{Tomasz Klimczuk}
\email{tomasz.klimczuk@pg.edu.pl}
\affiliation{Faculty of Applied Physics and Mathematics, Gdansk University of Technology, Narutowicza 11/12, 80-233 Gdańsk, Poland}
\affiliation{Advanced Materials Center, Gdansk University of Technology, Narutowicza 11/12, 80-233 Gdańsk, Poland}

\begin{abstract}

We present the results of a thorough investigation of the physical properties of EuAg$_4$Sb$_2$ single crystals using magnetization, heat capacity, and electrical resistivity measurements. High-quality single crystals, which crystallize in a trigonal structure with space group $R\bar{3}m$, were grown using a conventional flux method. Temperature-dependent magnetization measurements along different crystallographic orientations confirm two antiferromagnetic phase transitions around $T_{N1}$ = 10.5 K and $T_{N2}$ = 7.5 K. Isothermal magnetization data exhibit several metamagnetic transitions below these transition temperatures. Antiferromagnetic phase transitions in EuAg$_4$Sb$_2$ are further confirmed by two sharp peaks in the temperature-dependent heat capacity data at $T_{N1}$ and $T_{N2}$, which shift to the lower temperature in the presence of an external magnetic field. Our systematic heat capacity measurements utilizing a long-pulse and single-slope analysis technique allow us to detect a first-order phase transition in EuAg$_4$Sb$_2$ at 7.5 K. The temperature-dependent electrical resistivity data also manifest two features associated with magnetic order. The magnetoresistance exhibits a broad hump due to the field-induced metamagnetic transition. Remarkably, the magnetoresistance keeps increasing without showing any tendency to saturate as the applied magnetic field increases, and it reaches $\sim$20000\% at 1.6 K and 60 T. At high magnetic fields, several magnetic quantum oscillations are observed, indicating a complex Fermi surface. A large negative magnetoresistance of about -55\% is also observed near $T_{N1}$. Moreover, the $H$-$T$ phase diagram constructed using magnetization, heat capacity, and magnetotransport data indicates complex magnetic behavior in EuAg$_4$Sb$_2$.

\end{abstract}

	\maketitle
\section{INTRODUCTION}	

Even though there is a lack of single-ion anisotropy in divalent europium ($S$ = 7/2 and $L$ = 0), which results in negligible crystalline electric field (CEF) effects, its compounds often manifest intriguing magnetic phase transitions, leading to complex magnetic phase diagrams. Thus, Eu-based intermetallics have always been fascinating candidates for exploring complex magnetism  \cite{EuFe2As2_PRB_2005,EuFe2As2_PRB_2008,EuFe2As2_PRB_2019_Pressure_Neutron,EuFe2As2_SST_Ir_doped_2014,Eu_ternary_111_Pottgen,EuAuAs, EuRh2Si2,EuRhGe3}. Meanwhile, the discovery of chiral magnetic anomalies, the anomalous Hall effect, and the topological Hall effect in magnetic topological materials have drawn significant attention to the Eu-based compounds as they offer an ideal platform for studying the interplay between magnetism and band topology \cite{bernevig2022,tokura2019,Weyl_review,RAlGe,Mn3Sn}. EuCd$_2$As$_2$, EuMg$_2$Bi$_2$, Eu$T$As ($T$ = Au and Ag), and EuB$_6$ are the few examples of such magnetic topological materials  \cite{EuCd2As2_2020,EuCd2As2_2023, EuMg2Bi2_2021,EuMg2Bi2_2023,EuAuAs,EuAgAs, EuB6_2020,EuB6_2021}. Recently, the trigonal CaCu$_4$P$_2$-type Eu-based ternary pnictide has received significant attention, as this family of compounds reveals several interesting physical properties. For instance, two successive antiferromagnetic transitions below 15 K and several metamagnetic transitions under applied magnetic field are observed in  EuAg$_4$As$_2$ \cite{EuAg4As2_2021}. Interestingly, the system goes to an incommensurate noncollinear AFM state below 9 K, exhibiting an anomalous Hall effect. It also shows unusual magnetoresistance ($MR$) with large positive as well as negative values. The  \textit{MR} reaches 202\% at 2 K and -78\% around 10 K at 9 T. Similar magnetotransport behavior is observed in ferromagnetic EuCu$_4$As$_2$ \cite{EuCu4As2}. On the other hand, the non-magnetic members of this series, such as Sr$T_4Pn_2$ ($T$ = Ag and Cu, $Pn$ = As and Sb), CaCu$_4$As$_2$, and SrCu$_{4-x}$P$_2$, also feature several remarkable physical properties. For example, a phase transition associated with a structural distortion is observed in SrAg$_4$As$_2$, which also shows quantum oscillations associated with small Fermi pockets. Intriguingly, some of the compounds, such as SrAg$_4$Sb$_2$, CaCu$_4$As$_2$ and SrCu$_{4-x}$P$_2$, host nontrivial band topology, which results in the observation of large non-saturating magnetoresistance, quantum oscillations, and multilayer quantum Hall effect  \cite{SrAg4As2,SrAg4Sb2,SrCu4xP2,ACu4As2,CaCu4As2}.

Here we explore another compound, EuAg$_4$Sb$_2$, from the same family, which is the magnetic analogue of SrAg$_4$Sb$_2$ and has the potential to show complex magnetic behavior and topological phenomena. Earlier investigation revealed that EuAg$_4$Sb$_2$ crystallizes in CaCu$_4$P$_2$-type centrosymmetric trigonal structure with space group $R\bar{3}m$ (No. 166); it orders antiferromagnetically below 11 K and exhibits a metamagnetic transition at 0.24 T \cite{EuAg4Sb2}. However, there is no detailed investigation of the physical properties of single crystals. This report thoroughly investigates the physical properties of single crystals of EuAg$_4$Sb$_2$ along different crystallographic orientations using magnetic, thermodynamic, and magnetotransport measurements. Magnetic measurements reveal two antiferromagnetic phase transitions and multiple metamagnetic phase transitions. Temperature-dependent heat capacity and resistivity measurements further corroborate the magnetic phase transitions in EuAg$_4$Sb$_2$. Careful heat capacity measurements allow us to detect a first-order phase transition at 7.5 K. We have observed unusually large positive as well as negative magnetoresistance in EuAg$_4$Sb$_2$. The constructed magnetic phase diagram suggests a complex magnetic structure for EuAg$_4$Sb$_2$.

\section{EXPERIMENTAL DETAILS}

\begin{figure}
	\includegraphics[width=8.5cm, keepaspectratio]{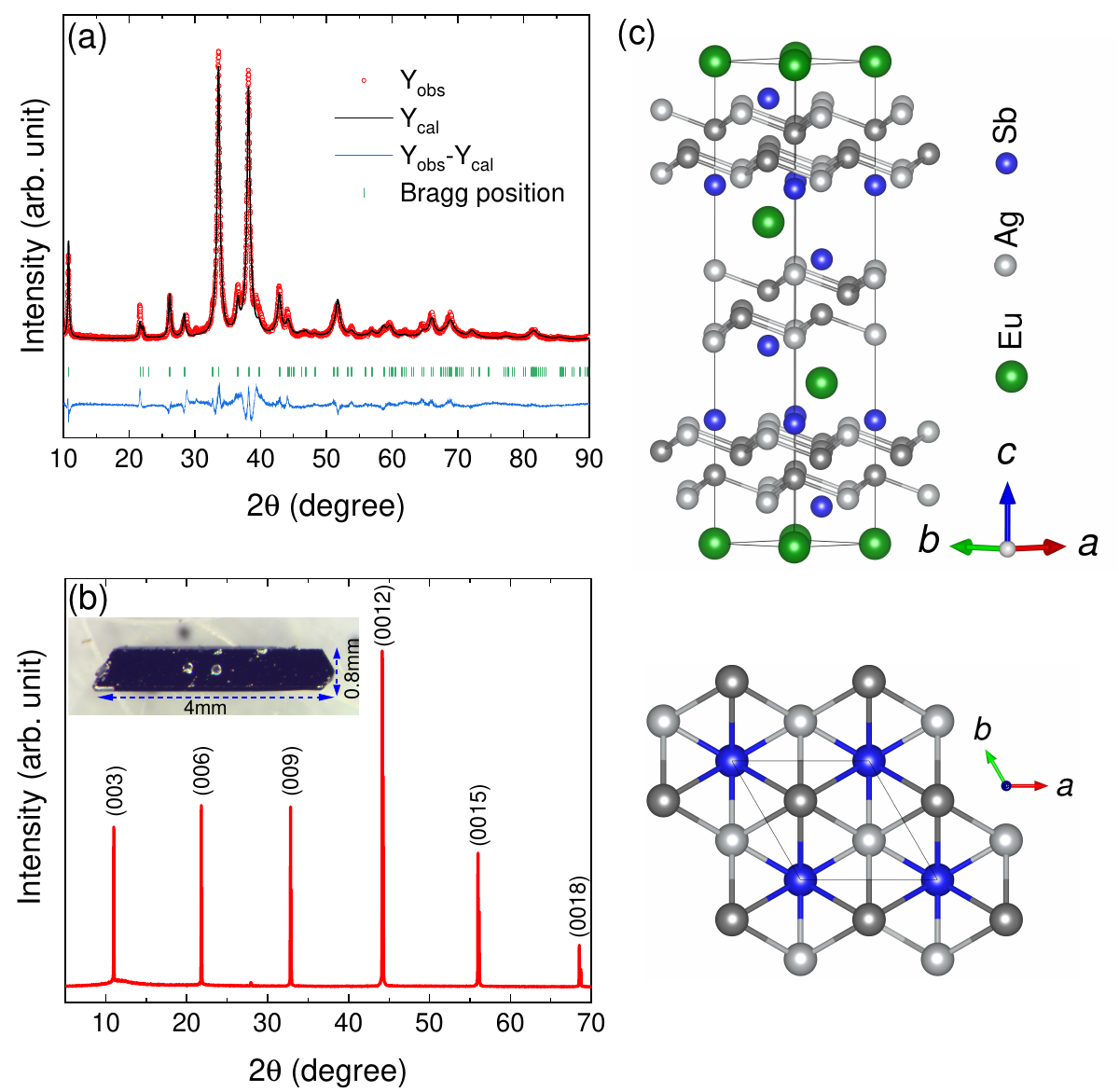}
	\caption{\label{XRD} (a) Powder XRD pattern of the crushed single crystals with the Rietveld refinement analysis. The red dots are the experimentally observed data, and the black line is the calculated XRD pattern from the analysis. The blue line represents the difference between the intensity of experimental and calculated data. Green vertical lines are the Bragg positions. (b) Single crystal XRD pattern. The inset shows an optical image of a single crystal. (c) A schematic diagram of the crystal structure of EuAg$_4$Sb$_2$. 
}
\end{figure}

The single crystals of EuAg$_4$Sb$_2$ were grown using Ag-Sb binary flux as described in Ref. \cite{SrAg4Sb2} for SrAg$_4$Sb$_2$. High purity elements Eu (99.9\%, Onyxmet), Ag (99.999\%, Alfa Aesar), and Sb (99.999\%, Alfa Aesar) were taken in 1:7:4 molar ratio. All the elements were mixed thoroughly and put into an alumina crucible. The crucible was sealed in a quartz ampule with partial argon pressure. The ampule was then heated to 1000$^\circ$C for 10 h and slowly cooled down to 600$^\circ$C at a rate of 1.5$^\circ$C/h. The ampule was taken out from the furnace at 600$^\circ$C and the crystals were separated from the flux by centrifuging. An optical image of a single crystal is displayed in the inset in Fig \ref{XRD}(b). The crystal structure was confirmed by x-ray diffraction (XRD) using a Bruker D2Phaser diffractometer with Cu K$_{\alpha1}$ radiation. The chemical composition was checked by energy dispersive x-ray spectroscopy (EDS) employing a FEI Quanta FEG 250 electron microscope. All the physical properties of the compound were measured in a Quantum Design EverCool-II Physical Property Measurement System (PPMS). Electrical resistivity measurements were performed using conventional four-probe techniques. Measurements of heat capacity were carried out using the short as well as long heat pulse method on the PPMS platform. Magnetic measurements were conducted using a vibrating sample magnetometer (VSM) attached to the PPMS. The high magnetic field data were taken at the pulsed-field facility of the National High Magnetic Field Laboratory (NHMFL, Los Alamos). We performed Proximity Detector Oscillator (PDO) measurements using a 10-turn pancake coil made of 50-gauge copper wire. The coil was connected to the PDO circuit with a resonant frequency of 26 MHz.  The technique is sensitive to the skin depth, and therefore can be used to probe changes in conductivity, including magnetic quantum oscillations \cite{PDO,PDO_PhysRevB}. The magnetoresistivity measurements were performed using standard four-probe technique. An AC current of 1 mA was applied at f = 100 kHz.

\section{Results and discussion}

\subsection{Crystal structure}

The powder and single crystal XRD patterns collected at room temperature are presented in Figs. \ref{XRD} (a) and \ref{XRD}(b), respectively. The powder XRD pattern was analyzed using the Rietveld refinement method in Fullprof software, which suggests EuAg$_4$Sb$_2$ crystallizes in trigonal structure with space group $R\bar{3}m$ (No. 166), which can be derived from the CaAl$_2$Si$_2$ \cite{CaAl2Si2} structure type by placing two additional silver atoms within the slab, creating [Ag$_4$Sb$_2$]$^{-2}$ layers. A schematic diagram of the crystal structure is shown in Fig. \ref{XRD}(c). Within each layer, the Ag1 and Ag2 atoms form buckled hexagons, which can be seen in the lower part of the figure. The estimated lattice parameters from the refinement are $a$ = 4.7189(3) \AA~ and $c$ = 24.661(3) \AA~, which agree well with the previous report \cite{EuAg4Sb2}. Fig \ref{XRD}(b) shows the single-crystal XRD pattern of EuAg$_4$Sb$_2$, which has extremely sharp reflections along (00$c$) directions, suggesting that the crystallographic $c$ axis is perpendicular to the flat plane of the single crystals. The sharp peaks indicate high crystalline quality of the grown single crystals. Additionally, EDS data acquired from various points on the single crystals confirms the expected stoichiometry.

\subsection{Magnetic Properties}
\begin{figure*}
	\includegraphics[width=17.8cm, keepaspectratio]{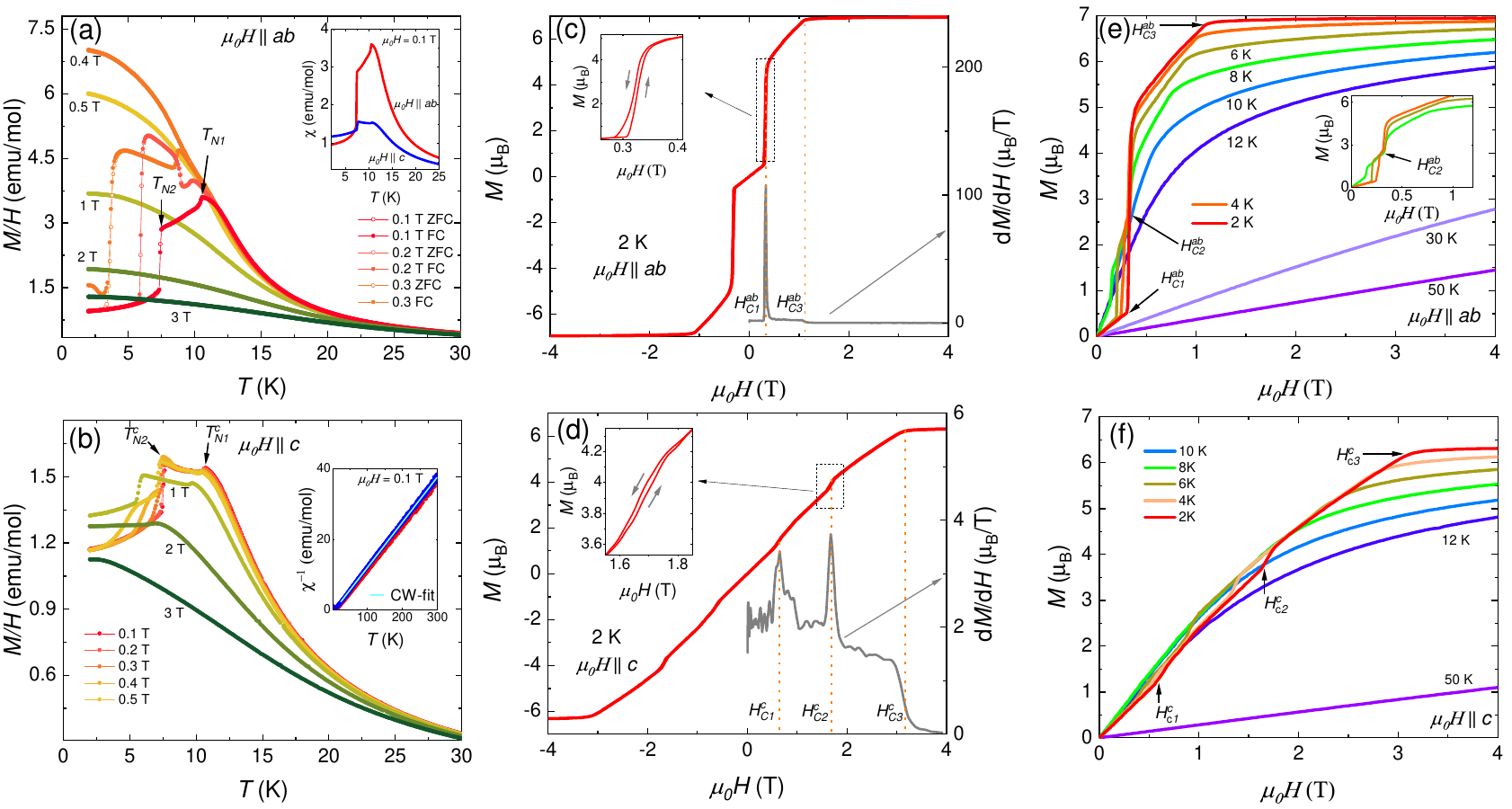}
	\caption{\label{MT} Magnetic properties of single-crystal EuAg$_4$Sb$_2$: (a) and (b) $M/H$ data as a function of temperature for various fields applied along the $H\parallel ab$ and $H\parallel c$ crystallographic orientations, respectively. The inset of (a) shows the magnetic susceptibility in the 2–25 K temperature range for $\mu_0H$ = 0.1 T along different crystallographic directions. Inverse susceptibility and CW fits up to room temperature are presented for both directions in the inset of (b). Isothermal magnetization for magnetic fields of up to $\pm$4 T at 2 K applied along $H\parallel ab$ (c) and $H\parallel c$ (d). Both the insets in (c) and (d) show the enlarged view of the $M(H)$ data, revealing a hysteresis loop. The right panels of (c) and (d) exhibit the first-order derivative of the $M(H)$ data, indicating the critical fields. Magnetization data at various temperatures up to 4 T for $H\parallel ab$ (e) and $H\parallel c$ (f). The inset of (e) indicates the critical field $H_{C2}^{ab}$.
}
\end{figure*}

Magnetization ($M$) data at various applied fields (H) in the zero-field cooled (ZFC) and field cooled (FC) configurations in the temperature range 2–30 K for $H\parallel ab$ and $H\parallel c$ crystallographic directions for a EuAg$_4$Sb$_2$ single crystal are presented in Figs. \ref{MT}(a) and \ref{MT}(b), respectively. The temperature-dependent $M/H$ data at 0.1 T along both crystallographic directions show a cusp and a kink around $T_{N1}$ = 10.6 K and $T_{N2}$ = 7.5 K, respectively, indicating antiferromagnetic (AFM) phase transitions. A similar temperature dependence of magnetic susceptibility ($\chi = M/H$) has also been observed previously in polycrystalline EuAg$_4$Sb$_2$ \cite{EuAg4Sb2}. As the applied field strength increases, the $T_{N1}$ and $T_{N2}$ shift to a lower temperature, a typical signature of an AFM \cite{EuMg2Bi2_2020}. Both the anomalies are smeared out for fields $\mu_0H \ge0.4$ T and $\mu_0H \ge 3$ T along $H\parallel ab$ and $H\parallel c$, respectively. Interestingly, a slight bifurcation is observed at low temperatures at $\mu_0H = 0.4$ T for $H\parallel ab$ in the ZFC and FC data, probably due to a field-induced metamagnetic (MM) transition. However, in the $H\parallel c$ direction, no irreversible behavior is seen between the ZFC and FC data. The magnetic susceptibility near the ordering increases significantly along $H\parallel ab$ compared to the $H\parallel c$ direction, as illustrated in the inset of Fig. \ref{MT}(a). On the other hand, at low temperatures (below $T_{N2}$), $\chi$ is larger along $H\parallel c$; it is nearly independent of temperature and does not drop significantly below the magnetic ordering temperature, as expected for a typical AFM. All these features suggest that Eu$^{2+}$ moments likely lie in the $ab$-plane and indicate a non-collinear incommensurate magnetic structure, as observed in the isostructural Eu-based compound EuAg$_4$As$_2$ \cite{EuAg4As2_2021,EuAg4As2_Neutron}. Moreover, at temperatures above magnetic order, the $\chi(T)$ data can be described using the Curie-Weiss (CW) law:

\begin{equation}
	\chi(T) = \dfrac{C}{T-\Theta_P}
\end{equation}

\noindent where $C$ is the Curie constant and $\Theta_P$ is the Curie temperature. The inset of Fig. \ref{MT}(b) shows a fit of the CW law to the $\chi^{-1}(T)$ data measured at 0.1 T in the temperature range 20-300 K. The estimated values of effective moment and Curie temperature are presented in Table \ref{table1}. The effective moment obtained is close to the theoretical value for the Eu$^{2+}$ ion ($g\sqrt{S(S+1)} = 7.94\mu_B$, $S = 7/2$ and $g = 2$). The fitted $\Theta_P$ is positive for both crystallographic directions, as observed in the polycrystalline sample, indicating a predominance of ferromagnetic exchange interactions in the paramagnetic state of EuAg$_4$Sb$_2$ \cite{EuAg4Sb2}.

\begin{table}
	\centering
	\caption {The estimated effective moment and Curie temperature along different crystallographic orientations are obtained from the CW fit.}
	\label{table1}
	\vskip .2cm
	\addtolength{\tabcolsep}{+20pt}
	\begin{tabular}{c c c }
		\hline
		\hline
		& $H\parallel ab$  &$H\parallel c$ \\[0.5ex]
		\hline
		$\mu_{eff} (\mu_B)$                   & 7.91(1)           & 7.86(1)  \\[1ex]
		$\Theta_P$(K)                         & 14.2(2)	          & 7.6(1)  \\[1ex]
		
		\hline
		\hline
	\end{tabular}
\end{table}

Figs. \ref{MT}(c) and \ref{MT}(d) show the magnetization data acquired at 2 K for fields of up to $\pm$4 T applied along the $H\parallel ab$ and $H\parallel c$ crystallographic orientations. As the magnetic field increases, the magnetization initially increases linearly, consistent with expectations for an AFM; but, around  $H_{C1}^{ab}$ = 0.33 T ($H\parallel ab$) and  $H_{C1}^{c}$ = 0.65 T ($H\parallel c$) there is a sudden jump in magnetization. For a field larger than $H_{C1}$, $M$ increases linearly until  $H_{C2}^{c}$ = 1.67 T for $H\parallel c$, where it shows another jump in the magnetization. With further increase of applied field strength, the magnetization saturates near  $H_{C3}^{ab}$ = 1.13 T and  $H_{C3}^{c}$ = 3.16 T. The saturation values are 6.3 $\mu_B$ ($H\parallel ab$) and 6.9 $\mu_B$ ($H\parallel c$), close to the theoretical value $gS\mu_B$ = 7 $\mu_B$ for the Eu$^{2+}$ ion. The derivative of $M$ as a function of $\mu_0H$, depicted in the right panel of Fig. \ref{MT}(c), reveals the critical fields where those magnetization jumps occur. Such jumps in magnetization are due to field-induced metamagnetic transitions. Interestingly, a hysteresis loop is seen in the metamagnetic transition corresponding to $H_{C2}^{ab}$ and $H_{C1}^c$ for increasing and decreasing fields, as illustrated in the inset of Figs. \ref{MT}(c) and \ref{MT}(d), respectively. The observed hysteresis around the metamagnetic transitions indicates that these phase transitions are likely first-order \cite{CaCo2As2}. We have also measured magnetization at various temperatures to get more insight into these transitions, as shown in Figs. \ref{MT}(e) and \ref{MT}(f). As the temperature increases, all the critical fields shift to lower values, and the observed anomalies disappear. Surprisingly, along $H\parallel ab$, an additional metamagnetic transition emerges at $H_{C2}^{ab}$ $\sim$ 0.3 T for $T \ge$ 4 K and persists up to 10 K. For $T >$ 30 K, which is well above the magnetic ordering temperature, the magnetization increases linearly as one anticipates for a paramagnetic state at relatively low magnetic fields. Notably, the behavior of $M(H)$ data closely resembles that of the isostructural compound EuAg$_4$As$_2$, which reveals an incommensurate, non-collinear AFM state at low temperatures, as confirmed by a neutron diffraction study \cite{EuAg4As2_2021,EuAg4As2_Neutron}. Therefore, a thorough investigation using neutron diffraction measurements is important to understand the complex magnetic structure of EuAg$_4$Sb$_2$. Nevertheless, similar behavior of the magnetization isotherms is also observed in Eu-based compounds like Eu$_3$Ni$_4$Ga$_4$ \cite{Eu3Ni4Ga4}, EuFe$_2$As$_2$ \cite{EuFe2As2_2009}, and EuCuAs \cite{EuCuAs}.

\subsection{Heat capacity}

\begin{figure}
	\includegraphics[width=8.2cm, keepaspectratio]{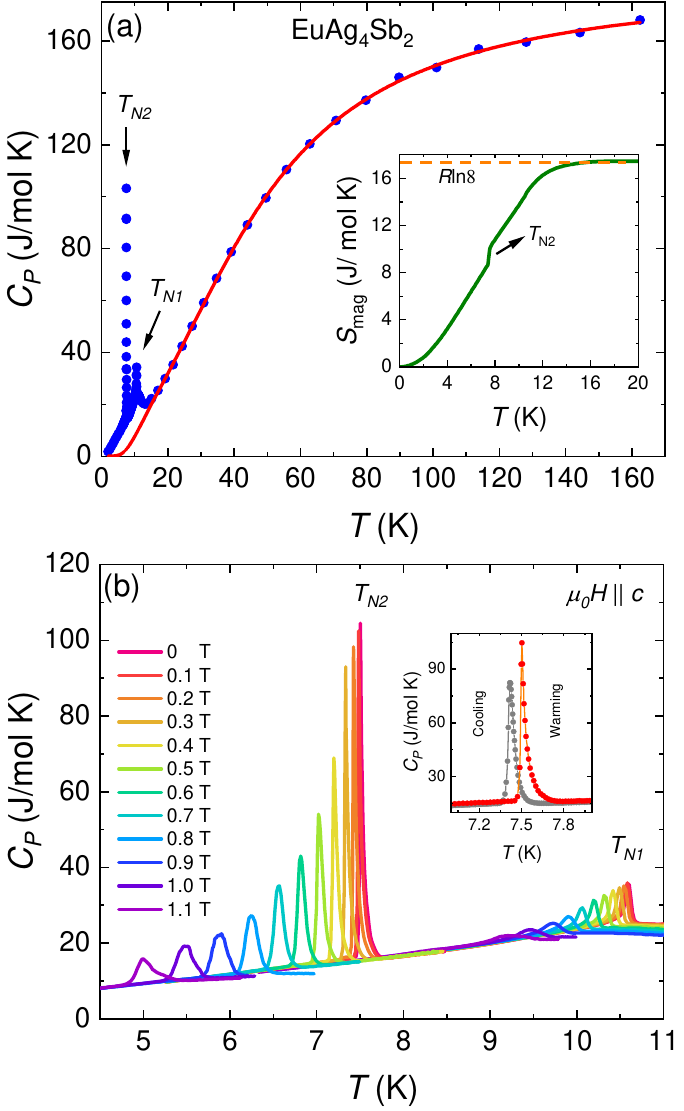}
	\caption{\label{HC}(a) The temperature dependence of heat capacity in the absence of external magnetic field measured in the temperature range 2-160 K. The red solid line represent the fitting of the data with the Eq. \ref{CP}. Inset shows the calculated magnetic entropy. (b) The evolution of the anomalies in the heat capacity under the applied magnetic field ranging from 0 to 1.1 T with a step of 0.1 T for the $H \parallel$ c. The inset depicts the zero-field $C_P$($T$) data obtained from the warming and cooling cycles near $T_{N2}$.}
\end{figure}

\begin{figure*}
	\includegraphics[width=16cm, keepaspectratio]{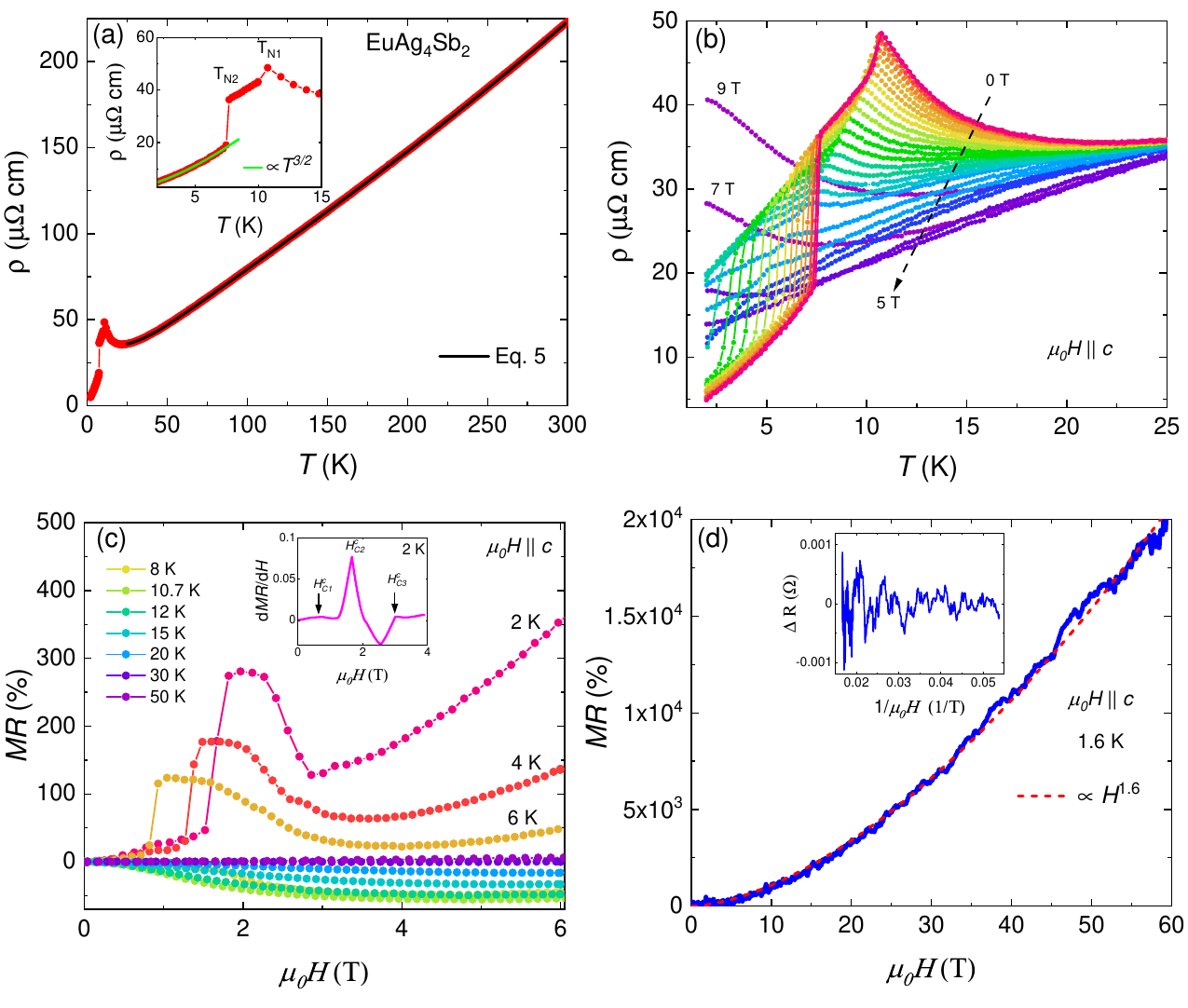}
	\caption{\label{RT}(a) Zero-field electrical resistivity over the temperature range 2 - 300 K. The inset shows an expanded view of $\rho$($T$) at low temperatures, highlighting the features due to the AFM transitions at $T_{N1}$ and $T_{N2}$. The solid green line represents the $T^{3/2}$ dependence of the electrical resistivity in the AFM ground state. (b) Temperature-dependent electrical resistivity under various applied fields 0 T $\le \mu_0H \le$ 9 T, where $H\parallel c$. (c) Magnetoresistance as a function of magnetic field applied along $H\parallel c$ at selected temperatures between 2 and 50 K. Inset shows the first derivative of  \textit{MR} as a function of the field, indicating the critical fields. (d) Field-dependent  \textit{MR} data up to 60 T at 1.6 K. The dashed red line demonstrates that  \textit{MR} follows the $H^{1.6}$ field dependence in the high-field region. The inset displays a clear quantum oscillation after a smooth background subtraction.}
\end{figure*}

Fig. \ref{HC}(a) shows the zero-field heat capacity ($C_P$) data at constant pressure as a function of temperature. At low temperatures, $C_P$($T$) exhibits two sharp anomalies at $T_{N1}$ = 10.5 K and $T_{N2}$ = 7.5 K as a manifestation of AFM transitions, as observed in the magnetic susceptibility data. The anomaly at $T_{N2}$ is extremely sharp, indicating a first-order phase transition, whereas the AFM transition at $T_{N1}$ is likely a second-order phase transition. A more detailed discussion about these transitions can be found later. The behavior of the $C_P$($T$) above the magnetic ordering temperature can be well replicated by considering the Debye ($C_D$) and Einstein ($C_E$)  models of specific heat, as combined in the equation  \cite{EuMg2Bi2_2020}

\begin{equation}
	C_P(T) = \gamma T+mC_{D}+(1-m)C_{E}~.
	\label{CP}
\end{equation}

\noindent The first term of the above equation is the electronic contribution to heat capacity. The $\gamma$ is known as the Sommerfeld coefficient. The weight factor $m$ establishes the proportionate contributions between $C_D$ and $C_E$ to the total heat capacity. We use the semi-empirical method used in Ref. \cite{EuMg2Bi2_2020} to treat the heat capacity of the related material EuMg$_2$Bi$_2$.  $C_D$ and $C_E$ are formulated as 
\begin{equation}
	C_{D}=9nR\left( \frac{T}{\Theta_D}\right)^3\int_{0}^{\Theta_D/T}\frac{x^4e^x}{(e^x-1)^2}dx~,
\end{equation}

\begin{equation}
	C_{E}=3nR\left( \frac{\Theta_E}{T}\right)^2\frac{e^{\Theta_E/T}}{(e^{\Theta_E/T}-1)^2}~.
\end{equation}

\noindent Fitting Eq. \ref{CP} to the heat capacity data yields $\gamma$ = 22.3 mJ/ mol K, $m$= 0.79, $\Theta_D$ = 208 K, and $\Theta_E$ = 50 K. The obtained fitting parameters are comparable to the isostructural compound EuCu$_4$As$_2$ \cite{EuCu4As2}. To get a rough estimation of magnetic entropy, we have extrapolated the fitting to 2 K and subtracted it from the experimental data to obtain the magnetic component ($C_{mag}$) of heat capacity. The inset of Fig. \ref{HC}(a) shows the magnetic entropy computed using the expression, $S_{mag} = \int \frac{C_{mag}}{T}dT$. The magnetic entropy at $T_{N1}$ is about 14.5 J/mol K, and it increases further as temperature increases and saturates around 16 K to an expected value of $S_{mag} = Rln(2S+1)$ = $Rln8$ for the Eu$^{2+}$ ion.

All the low-temperature heat capacity data presented in Fig. \ref{HC} has been measured using the long heat pulse technique, and the data extracted from the single-slope analysis method in Quantum Design PPMS, which is one of the most effective ways to identify a first-order phase transition \cite{LongHCPulse}. The observed peak at $T_{N2}$ is extremely sharp, having a full width at half maximum (FWHM) of less than 0.1 K. In addition, a thermal hysteresis of around 0.16 K appears in the warming and cooling cycle of the $C_P$($T$) data, as shown in the inset of Fig. \ref{HC}(b). Remarkably, a discontinuity in temperature-dependent magnetic entropy at $T_{N2}$ can be seen in the inset of Fig. \ref{HC}(a). All of these observations support the first-order nature of the AFM transition, probably attributed to the reconstruction of the AFM structure below $T_{N2}$ \cite{RTM2Al20,guillou2018non}.  Fig. \ref{HC}(b) depicts the $C_P$($T$) at low temperatures for various applied fields. Both features shifted to lower temperatures, with peaks broadening as the magnetic field intensity increased, which is a typical characteristic of an AFM.

\begin{figure}
	\includegraphics[width=8.6cm, keepaspectratio]{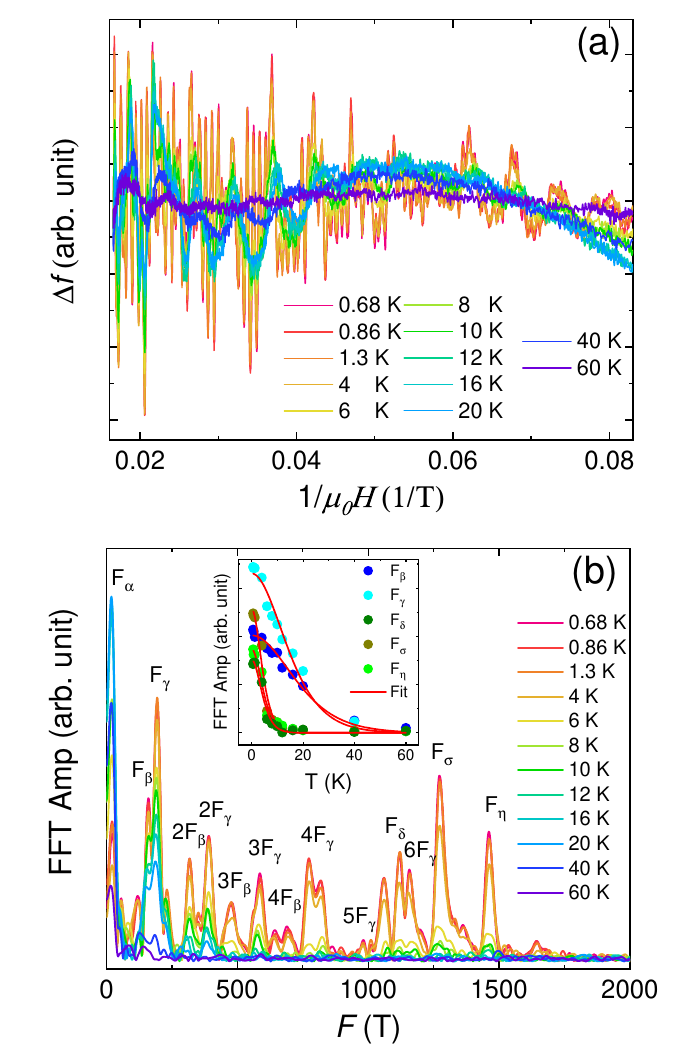}
	\caption{\label{SdH} (a) PDO data as a function of $1/\mu_0H$ after smooth background subtraction measured at 12 T window along $H\parallel c$. (b) The frequency-dependent FFT signals measured at various temperatures. The inset shows the fitting of the FFT amplitude to Eq. \ref{LK}.}
\end{figure}

\subsection{Magnetotransport}

\begin{table*}
	\centering
	\caption {The parameters obtained from the quantum oscillation.}
	\label{table2}
	\vskip .2cm
	\addtolength{\tabcolsep}{+12pt}
	\begin{tabular}{c c c c c c c}
		\hline
		\hline
		                      & $F_\alpha$  & $F_\beta$ & $F_\gamma$ &$F_\delta$ &$F_\sigma$ &$F_\eta$ \\[0.5ex]
		\hline
		                               
		Frequency (T)              & 22          & 160           & 193           & 1120          & 1273           & 1460            \\[1ex]
		$m^*$ ($m_e$)              & --          & 0.16 $\pm$ 0.01 & 0.19$\pm$ 0.01  & 0.67 $\pm$ 0.05 & 0.68 $\pm$ 0.05  & 0.64 $\pm$ 0.04  \\[1ex]
		$A_F$ (nm$^{-2}$)          & 0.21        & 1.53          & 1.84          & 10.70         & 12.16          & 13.94             \\[1ex]
		$k_F$ (nm$^{-1}$)          & 0.26        & 0.70          & 0.77          & 1.84          & 1.97           & 2.11               \\[1ex]
		$v_F$ (10$^6$ms$^{-1}$)    & --          & 0.51          & 0.47          & 0.32          & 0.34           & 0.38               \\[1ex]

		\hline
		\hline
	\end{tabular}
\end{table*}

The electrical resistivity $\rho$($T$) as a function of temperature measured in zero magnetic field for current along the $ab$-plane in the EuAg$_4$Sb$_2$  single crystal is shown in Fig. \ref{RT} (a). $\rho$($T$) decreases monotonically with decreasing temperature in a metallic fashion down to 25 K, then it starts to increase and exhibits a peak at 10.7 K and a kink at 7.5 K as the system goes through two antiferromagnetic phase transitions at these temperatures. Remarkably, the behavior of electrical resistivity at low temperatures mimics the magnetic susceptibility data seen in Fig. \ref{MT}(a). The residual resistivity ratio [RRR =$\rho$(300 K)/$\rho$(2 K) ] $\sim$ 50, is very high, suggesting the very good quality of the as-grown single crystals. A similar RRR value is also found in the nonmagnetic analogue compound SrAg$_4$Sb$_2$ \cite{SrAg4Sb2}. The behavior of $\rho$($T$) above the magnetic ordering temperature can be well explained by considering electron-phonon ($ep$) and electron-electron ($ee$) scattering as described by the expression below \cite{EP_Int}

\begin{equation}
\label{BG}
\small
\rho_{xx} (T)= \rho_0 + 4R\left( \frac{T}{\Theta_R}\right)^5\int_{0}^{\Theta_R/T}\frac{x^5}{(e^x-1)(1-e^{-x})}dx +aT^2,
\end{equation}

\noindent where the first term is the temperature-independent residual resistivity, the second term is the Bloch-Gruneisen (BG) scattering model, which accounts for $ep$ scattering, and the last term is the contribution from the $ee$ scattering. The fit parameters are $\rho_0$ = 34 $\mu \Omega$cm, $R$ = 404  $\mu \Omega$cm, $\Theta_R$ = 202 K, and $a$ = 4.7$\times$10$^{-4}$ $\mu \Omega$cm K$^{-2}$. The Debye temperature obtained is close to that estimated from the heat capacity data. Moreover, the temperature-dependent electrical resistivity below the magnetic phase transition follows a $T^{3/2}$ dependence, as found in several AFM \cite{GdCu6, NiS2-xSex}.

In order to determine the effect of magnetic field on the electrical resistivity, we conducted a systematic field-dependent measurement in the temperature range 2–25 K with the field applied along $H\parallel c$, as displayed in Fig. \ref{RT}(b). As field strength increases, the ordering temperatures shift to lower values, and both anomalies gradually disappear, in agreement with magnetic susceptibility data. At $\mu_0H$ = 3 T, all resistivity peaks disappear, suggesting metallic behavior. Interestingly, there is a large positive magnetoresistance at low temperatures for $\mu_0H \geq$ 4 T; at 2 K and 9 T, the resistivity reaches 40.6 $\mu \Omega$cm. Consequently, we measured  \textit{MR} at several temperatures with the field applied along the $c$-axis, as shown in Fig. \ref{RT}(c). \textit{MR} is calculated using the formula $\textit{MR} =\dfrac{\rho(H)-\rho(0)}{\rho(0)}\times 100\%$, where $\rho(0)$ and $\rho(H)$  are the resistivity values in the absence and presence of the external magnetic field. The  \textit{MR} measured at 2 K grows slowly with increasing field, but at about 1.6 T, there is a significant jump in  \textit{MR}, resulting in a broad hump which shifts to lower fields as the temperature increases, in accordance with the $M(H)$ data. The abrupt increase in  \textit{MR} can be attributed to increased magnetic scattering due to metamagnetic transitions. A significant additional contribution to the resistivity on crossing the boundary of the antiferromagnetic phase is likely to come from the rearrangement of the Fermi surface as the magnetic unit cell changes, resulting in an alteration of the density of states available for scattering close to the Fermi energy \cite{NdB}. The d\textit{MR}/d$H$ vs $\mu_0H$ data, as presented in the inset of Fig. \ref{RT}(c), exhibit three peaks, which indicate three critical fields associated with the metamagnetic transitions observed in the magnetization data.  \textit{MR} at low temperatures continues to increase without showing any saturation. For $\mu_0H>$ 3, all the spins are ferromagnetically aligned in the field direction, as indicated by the magnetization; thus, one may anticipate saturation or a decrease in the  \textit{MR} at high applied fields. Nevertheless, as the temperature increases, the  \textit{MR} starts to decrease, and near the ordering temperature, it becomes negative due to the suppression of magnetic scattering. At $T_{N1}$, it shows a maximum negative  \textit{MR} of about -55\%. Once the temperature increases beyond the ordering temperature, the  \textit{MR} starts to increase again. However, at temperatures well above magnetic order,  \textit{MR} is nearly independent of the external magnetic field. Such behavior of  \textit{MR} near the ordering temperature is primarily due to the response of magnetic scattering by the external field, as observed in similar Eu-based compounds, EuAg$_4$As$_2$ and EuCu$_4$As$_2$ \cite{EuAg4As2_2021,EuCu4As2}. In order to investigate the high-field behavior of the sample, additional measurements up to 60 T were performed at 1.6 K. The  \textit{MR} continues to increase even at the highest field and reaches a very large value of $\sim$ 20000\%, as shown in Fig. \ref{RT}(d). Moreover,  \textit{MR} follows a $H^{1.6}$ field dependence, unlike the semi-classical $H^2$ dependence \cite{Niu_2022}. A large  \textit{MR} is also observed in a similar compound, EuCu$_4$As$_2$, and its origin has been attributed to a nontrivial band structure \cite{EuCu4As2}. Likewise, nonmagnetic SrAg$_4$Sb$_2$ exhibits unusually high  \textit{MR} due to its nontrivial topological states \cite{SrAg4Sb2}. Thus, the large non-saturating  \textit{MR} and its deviation from the quadratic field dependence in EuAg$_4$Sb$_2$ may result from a nontrivial band structure. The high-field  \textit{MR} data also indicates the presence of quantum oscillations with multiple frequencies, suggesting a complex Fermi surface.

Fig. \ref{SdH}(a) shows quantum oscillations after background subtraction using a polynomial function in the contactless electrical resistivity data of EuAg$_4$Sb$_2$ measured using the proximity detector oscillator technique at high magnetic fields (12 T $\le \mu_0H \le$ 60 T) and in the temperature range 0.68-60 K. The PDO measurement detects Shubnikov-de Haas (SdH) oscillations in the electrical resistivity, via the change in resonant frequencies ($f$) of the oscillator circuit due to a change in the skin depth of the sample. The fast Fourier transform (FFT), as shown in Fig. \ref{SdH}(b), reveals several fundamental frequencies and their harmonics. The cyclotron effective mass can be calculated by fitting the amplitude FFT signal to the temperature-dependent part of the Lifshitz-Kosevich formula \cite{shoenberg}

\begin{equation}
\label{LK}
    A(T, B_m) \propto \dfrac{14.69m^*T/B_m}{sinh(14.69m^*T/B_m)}
\end{equation}

\noindent where $B_m$ is the inverse-field midpoint of the field window used for the FFT, defined as

\begin{equation}
    B_m = \left[\dfrac{1}{2}\left(\dfrac{1}{B_1}+\dfrac{1}{B_u}\right)\right]^{-1}
\end{equation}

\noindent Here, $B_l$ and $B_u$ are the lower and upper field limits of the window, respectively. The estimated value of the effective mass and its corresponding frequency are presented in Table \ref{table2}. Frequency $F_{\alpha}$ exhibits unusual temperature dependence, which limits the estimation of effective mass and is subject to further study. Using the Onsager relation $F = (\hbar/2\pi e)A_F$, we have calculated the cross-sectional areas of the Fermi surface ($A_F$) perpendicular to the applied magnetic field. Further, under the assumption of a circular Fermi-surface cross-section, we have estimated the Fermi wave vectors ($k_F$) and the Fermi velocities ($v_F$) as presented in Table  \ref{table2} using the expressions  $k_F = \sqrt{2eF/\hbar}$ and $v_F = \hbar k_F/m^*$, respectively, where $\hbar$ is the reduced Planck’s constant and $e$ is the magnitude of electron charge.

\subsection{$H$-$T$ phase diagram}

\begin{figure}
	\includegraphics[width=8.0cm, keepaspectratio]{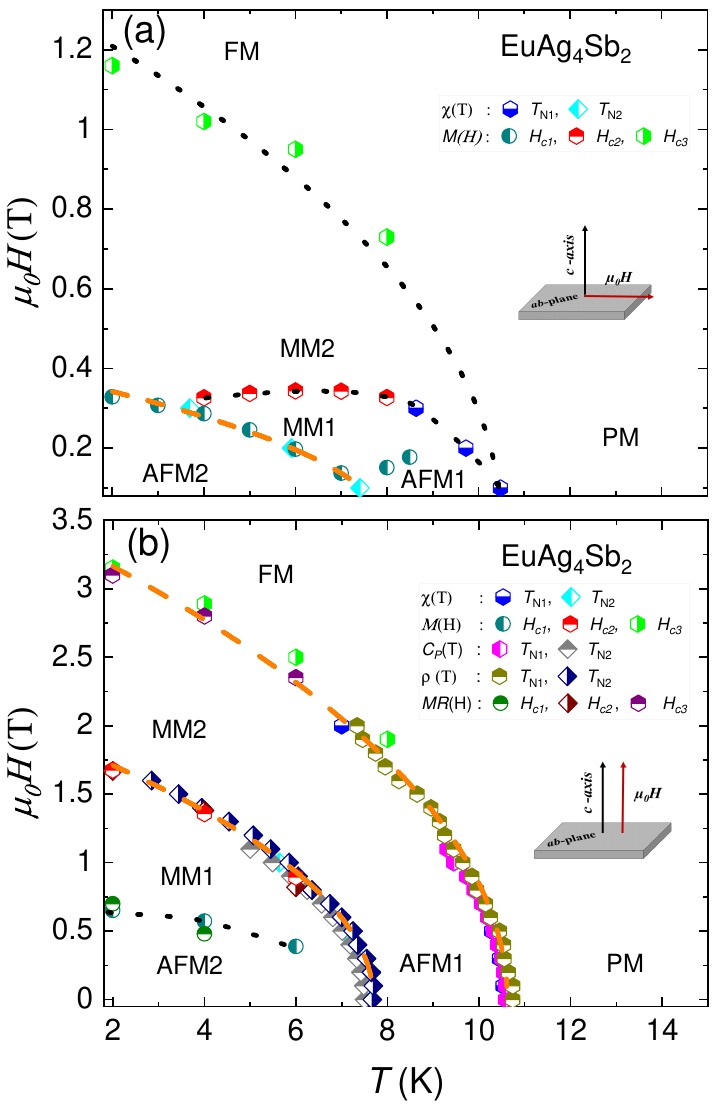}
	\caption{\label{HT} Magnetic phase diagram of EuAg$_4$Sb$_2$ single crystal for (a) $H\parallel ab$ and (b) $H\parallel c$ orientations, constructed using magnetization, heat capacity, and magnetotransport data. Black dot lines are the hand drawing for visual guides. Orange dash lines originate from the molecular field theory expression given in the text. }
\end{figure}

To get an overall picture of the magnetic behavior of EuAg$_4$Sb$_2$ at low temperatures, we have constructed an $H$-$T$ phase diagram based on the magnetic, heat capacity, and magnetotransport measurements for the field orientations $H\parallel ab$ and $H\parallel c$, as presented in Figs. \ref{HT}(a) and \ref{HT}(b), respectively. The overall $H$-$T$ phase diagram can be divided into six domains. As the temperature decreases from room temperature, the system enters the first antiferromagnetic (AFM1) state from the paramagnetic (PM) state around 10.5 K. With a further decrease in temperature, the system exhibits a first-order AFM phase transition, as confirmed by the heat capacity data. The region below this transition is labeled as AFM2. Both ordering temperatures shift to lower temperatures with increasing magnetic field, forming a distinct phase diagram boundary in accordance with the molecular field theory expression $H = H_0[1-T_N(H)/T_N(H=0)]^{1/2}$, where $H_0$ is the critical field that destroys the antiferromagnetic transitions at 0 K \cite{EuCo2P2}. The estimated values of $\mu_0H_0$ for $H\parallel ab$ are 0.39 T ($T_{N2}$), whereas for $H\parallel c$, $\mu_0H_0$ = 3.5 T ($T_{N1}$) and 1.98 T ($T_{N2}$). When a magnetic field of $H_{C1}$ is applied to the system within the AFM2 regime, the system undergoes a metamagnetic transition (MM1). For $H\parallel ab$, the phase boundaries MM1 and AFM2 overlap with each other. However, for $H\parallel c$, MM1 and AFM2 create two separate regions. Another metamagnetic transition (MM2) occurs when the magnetic field hits $H_{C2}$. These MM transitions are most likely the results of spin-flop transitions \cite{CaCo2As2}. At low temperatures (below 6 K), the AFM2 and MM2 phase boundaries overlap in the case of $H\parallel c$. As the applied magnetic field increases further, the system undergoes another transition at $H_{C3}$, at which point all spins align along the field direction; it behaves like a ferromagnet (FM). Interestingly, for $H\parallel c$, the phase boundary for $H_{C3}$ at higher fields almost overlaps with the first AFM transition. However, we could not clearly distinguish between PM and FM states based on the available data.

\section{Conclusion}

We have studied the magnetic, thermodynamic, and electrical transport properties of high-quality single-crystal EuAg$_4$Sb$_2$ grown using Ag-Sb binary flux. The XRD data confirm that EuAg$_4$Sb$_2$ crystallizes in a trigonal structure. Our comprehensive analysis of temperature-dependent magnetization, electrical resistivity, and heat capacity data reveals that EuAg$_4$Sb$_2$ orders antiferromagnetically at 10.5 K and 7.5 K. The transition at 7.5 K is identified as a first-order phase transition using systematic heat capacity measurements performed with the long heat pulse technique. Several metamagnetic transitions are observed in the isothermal magnetization data. Further investigation using neutron diffraction measurements is required to fully understand the complex magnetic structure of EuAg$_4$Sb$_2$. Interestingly,  \textit{MR} displays non-saturation behavior at low temperatures, which reaches a very large value of $\sim$20000\% at 1.6 K and 60 T. Moreover, as the temperature rises, the  \textit{MR} decreases and becomes negative. Near the first AFM phase transition, a large negative value of about 55\% is observed. In addition, the field-dependent magnetoresistance shows a hump as a characteristic of metamagnetic transitions. At high magnetic fields, several magnetic quantum oscillations are observed, indicating a complex Fermi surface. Our findings on EuAg$_4$Sb$_2$ indicate that it is an antiferromagnetic system with a complex magnetic structure and a potential topological material. 

\section{ACKNOWLEDGMENTS}

The work at Gdansk University of Technology was supported by the National Science Center (Poland), Grant No. 2022/45/B/ST5/03916. A portion of this work was performed at the National High Magnetic Field Laboratory, which is supported by National Science Foundation Cooperative Agreement No. DMR-2128556, the US Department of Energy (DoE)  and the State of Florida. JB and JS acknowledge support from the DoE BES FWP “Science of 100 T”, which permitted development of some of the high-field techniques used in the paper

\bibliography{EuAg4Sb2}
	
\end{document}